\documentstyle[aps,prd,a4,preprint]{revtex}
\input feynman
\input epsf

\newlength{\fermilength}
\newlength{\gluonfourlength}
\newlength{\templength}

\newcommand{\be}{\begin{equation}}
\newcommand{\ee}{\end{equation}}
\newcommand{\bea}{\begin{eqnarray}}
\newcommand{\eea}{\end{eqnarray}}
\newcommand{\eps}{\varepsilon}
\newcommand{\dfx}[1]{d^4\!x_{#1}\,}
\newcommand{\Adash}[0]{/\hspace{-0.7em}A}

\newcommand{\D}[0]{\displaystyle}

\newcommand{\scs}[0]{\scriptsize}
\newcommand{\ei}[0]{\eps_i}
\newcommand{\ef}[0]{\eps_f}
\newcommand{\eq}[0]{\eps_q}
\newcommand{\eqp}[0]{\eps_{q'}}
\newcommand{\eip}[0]{\eps_{i'}}
\newcommand{\efp}[0]{\eps_{f'}}
\newcommand{\oms}[0]{\Omega_m^\Sigma}
\newcommand{\omsp}[0]{\Omega_{m'}^{\Sigma'}}
\newcommand{\ket}[1]{|#1\rangle}

\newcommand{\half}[0]{\frac{1}{2}\,}
\newcommand{\alphas}[0]{\alpha_{\rm s}}

\begin{document}
\title{Hadron masses in cavity quantum chromodynamics to order \boldmath $\alpha_{\rm s}^2$}
\author{M.\ Schumann\thanks{Present Address: Institut f\"ur Theoretische Physik,
Universit\"at Basel, 4056 Basel, Switzerland,
Email: Marc.Schumann@unibas.ch}, R.J.\ Lindebaum 
	and R.D.\ Viollier}
\address{Institute of Theoretical Physics and Astrophysics, 
University of Cape Town, Rondebosch 7701, South Africa}

\date{February 29, 2000}
\maketitle

\begin{abstract}
The non-divergent diagrams describing two-gluon exchange and
annihilation between quarks and antiquarks are calculated in the
Feynman gauge, based on quantum chromodynamics in a spherical cavity.
Using the experimental $N$, $\Delta$, $\Omega$, and $\rho$ masses to
fit the free parameters of the M.I.T.\ bag model, the predicted states
agree very well with the observed low-lying hadrons.  As expected, the
two-gluon annihilation graphs lift the degeneracy of the $\pi$ and
$\eta$, while the $\rho$ and $\omega$ remain degenerate. Diagonalizing
the $\eta - \eta'$ subspace Hamiltonian yields a very good
value for the mass of the $\eta$ meson.
\end{abstract}

\section{Introduction}
Cavity quantum chromodynamics, i.e.\ 
quantum chromodynamics with field operators obeying the linear
boundary conditions of the M.I.T.\ bag model on a statics sphere
(CQCD) \cite{Chodos74a,Chodos74b,DeGrand75}, is a consistent
relativistic quantum field theory on its own. Indeed, CQCD may be
expanded perturbatively \cite{Buser88}, and the diverging Feynman
graphs have been shown to be renormalizable 
\cite{Page93,Schreiber92b,Stoddart90,Cuthbert93}, e.g.\ in the
$\overline{\rm MS}$ scheme. Taking into account the quadratic boundary
condition of the M.I.T.\ bag model, the properties of the low-energy
hadrons have been calculated successfully to order $\alphas$ 
\cite{Chodos74a,Chodos74b,DeGrand75}.

However, this good agreement has been obtained by neglecting the
self-energies of the quarks. More
recently, the self-energies have been calculated, but unfortunately they
turned out to be quite large
\cite{Schreiber92b,Stoddart90,Cuthbert93}, thus spoiling the good
agreement to order $\alphas$. One has therefore argued that the
self-energies should perhaps be discarded, because the boundary
conditions already account for at least part of them.

Moreover, it has been pointed out that the perturbative expansion of CQCD in
terms of a power series in the large strong coupling constant may be
ill-defined. However, one can argue that the actual value of $\alphas$
is immaterial, as long as the spectrum fits, order by order, the same
hadronic states, $N$, $\Delta$, $\Omega$, and $\rho$, thus fixing the
free parameters of the model. Furthermore, it seems obvious that
perturbative expansion in terms of cavity modes, rather than plane
waves, is a much better starting point for the description of
finite size hadrons.

In spite of all these arguments, it is surprising to note that there
was in the recent past little enthusiasm to develop CQCD beyond first
order to check whether these expectations are actually fulfilled or
not. It is, therefore, the purpose of this paper to clarify this
issue, by calculating all non-divergent graphs to order $\alphas^2$ in
CQCD that can be written in the form of two-body operators, and to compare 
the resultant low-lying hadron spectrum with the
experimental data. We are concentrating only on the two-body interactions in
this paper. A calculation of the nondivergent three-body interactions for
massless quarks has been performed showing that they are  of much less 
importance than the two-body interactions to order $\alphas^2$. These
interactions leading to a three-body force are not included in the following.


\section{Second order energy shift}
Using the symmetric form of the Gell-Mann and Low theorem due to Sucher 
\cite{Sucher57}, 
we can write the energy shift of an eigenstate $|\phi_k\rangle$ 
of the unperturbed Hamiltonian as
\be
E_k-E^0_k=\lim_{{\eta\to 1}\atop {\eps\to 0_+}}
\frac{i\eps}{2}\frac{\partial\langle\hat\phi_k|S_\eta^\eps|\hat\phi_k\rangle_c/\partial\eta}
{\langle\hat\phi_k|S_\eta^\eps|\hat\phi_k\rangle_c},
\ee
where the subscript $c$ indicates that only connected diagrams are included. 
The adiabatic $S$-matrix may be expanded as
\be
S_\eta^\eps=1+\sum_{n=1}^\infty S_\eta^{\eps(n)}
\ee
with
\be
S_\eta^{\eps(n)}=\frac{(i\eta)^n}{n!}\int_{-\infty}^\infty dt_1\dots
\int_{-\infty}^\infty dt_n\,T\left[\hat H_{\rm int}^\eps(t_1)\cdots
\hat H_{\rm int}^\eps(t_n)\right].
\ee
Here, the parameter $\eps>0$ introduces the adiabatic switching on
of the interaction Hamiltonian $\hat H_{\rm int}^\eps(t)$ in the Dirac picture.
Expanding $\Delta E$ up to second order in $\alpha_{\rm s}$, 
we find that the only 
non-divergent terms contributing to $\Delta E$ are 
$\langle S^{\eps(2)}\rangle_c$ and
$\langle S^{\eps(4)}\rangle_c$, and thus $\Delta E$ reads to that order
\be
\Delta E=\lim_{\eps\to 0_+}
i\eps \left[ \langle S^{(2)}\rangle_c + 2\,\langle S^{(4)}\rangle_c
\right].
\label{deltaE}
\ee
The first term gives rise to the well-known one-gluon exchange and annihilation
graphs \cite{Buser88}, while the second contains the two-gluon exchange and 
annihilation graphs which we would like to evaluate in this paper.

Stoddart {\it et al.} 
\cite{Stoddart88}
have calculated the two-body operators in second order including 
six of the 24 possible time-orderings for both the two-gluon exchange
and annihilation graphs. Here we take into
account all time-orderings, using the results of ref.~\cite{Stoddart88} 
as a check. In fact, all the results of ref.~\cite{Stoddart88}  can
be reproduced with our method.

As an example, let us calculate the `straight' two-gluon exchange graph
(Fig.\ \ref{Feynman}(s2gx)).  The energy shift due to this diagram is 
given by
\ifpreprintsty
{\be
\begin{array}{rcl}
\D\Delta E &=
&\D \lim \limits_{\eps\rightarrow 0_+} 2i\eps 
 \int \dfx{1}\dfx{2}\dfx{3}\dfx{4}
 e^{-\eps(|t_1|+|t_2|+|t_3|+|t_4|)}\\*[3mm]
&&\D
\hspace{-1cm}\begin{array}{l}
 \D {} \times \left\langle \hat{\phi_k} \left|\,T\!\left[
\left(\hat{\bar\psi}g\frac{\lambda_a}{2}\hat{\Adash}_a\hat\psi\right)_{\!\!x_1}
\!\!
\left(\hat{\bar\psi}g\frac{\lambda_b}{2}\hat{\Adash}_b\hat\psi\right)_{\!\!x_2}
\!\!
\left(\hat{\bar\psi}g\frac{\lambda_c}{2}\hat{\Adash}_c\hat\psi\right)_{\!\!x_3}
\!\!
\left(\hat{\bar\psi}g\frac{\lambda_d}{2}\hat{\Adash}_d\hat\psi\right)_{\!\!x_4}
 \right]\right|\hat{\phi_k}\right\rangle\, . \\
\D \phantom{\times}
\rule{7.3em}{0pt}\rule{0.5pt}{4mm}
	 \rule{11.4em}{0.5pt} \rule{0.5pt}{4mm}
	 \raisebox{2mm} {\hspace{-10.8em} $\rule{0.5pt}{2mm}
			   \rule{2.4em}{0.5pt}  \rule{0.5pt}{2mm} 
			  \rule{9.0em}{0pt} \rule{0.5pt}{2mm}
				\rule{2.4em}{0.5pt} \rule{0.5pt}{2mm}$}
	 \raisebox{-2mm} {\hspace{-9.5em} 
		  $\rule{0.5pt}{6mm}\rule{11.4em}{0.5pt}\rule{0.5pt}{6mm}
			   $} 
       \end{array}
\end{array}
\ee}%
\else
{\be
\begin{array}{rcl}
\D\Delta E &=
&\D \lim \limits_{\eps\rightarrow 0_+} 2i\eps 
 \int \dfx{1}\dfx{2}\dfx{3}\dfx{4}
 e^{-\eps(|t_1|+|t_2|+|t_3|+|t_4|)}\\*[3mm]
&&\D
 \begin{array}{l}
 \D {}\times \left\langle \hat{\phi_k} \left|\,T\!\left[
\left(\hat{\bar\psi}g\frac{\lambda_a}{2}\hat{\Adash}_a\hat\psi\right)_{\!\!x_1}
\!\!
\left(\hat{\bar\psi}g\frac{\lambda_b}{2}\hat{\Adash}_b\hat\psi\right)_{\!\!x_2}
\!\!
\left(\hat{\bar\psi}g\frac{\lambda_c}{2}\hat{\Adash}_c\hat\psi\right)_{\!\!x_3}
\!\!
\left(\hat{\bar\psi}g\frac{\lambda_d}{2}\hat{\Adash}_d\hat\psi\right)_{\!\!x_4}
 \right]\right|\hat{\phi_k}\right\rangle\ . \\
\D \phantom{{}\times}
\rule{7.8em}{0pt}\rule{0.5pt}{4mm}
     \rule{12.0em}{0.5pt} \rule{0.5pt}{4mm}
     \raisebox{2mm} {\hspace{-11.4em} $\rule{0.5pt}{2mm}
               \rule{2.7em}{0.5pt}  \rule{0.5pt}{2mm} 
              \rule{9.2em}{0pt} \rule{0.5pt}{2mm}
                \rule{2.7em}{0.5pt} \rule{0.5pt}{2mm}$}
     \raisebox{-2mm} {\hspace{-10.1em} 
          $\rule{0.5pt}{6mm}\rule{12.0em}{0.5pt}\rule{0.5pt}{6mm}
               $} 
       \end{array}
\end{array}
\ee}%
\fi
The time integrals are readily evaluated and the space integrals combine into
quark-gluon vertex integrals defined by
\be
Q_{nn'}^{m\Sigma}=i\int d^3x\,\bar{u}_n(\vec x)\,\gamma_\mu\,u_{n'}(\vec x)\,
a_{m\Sigma}^\mu(\vec x),
\ee 
where $u_n(\vec x)$ and $a_{m\Sigma}^\mu(\vec x)$ are the quark and gluon
cavity modes, respectively \cite{OConnor96}.
The result is 
\bea
 \Delta E&=& 
 -g^4
 \left\langle \hat{\phi_k}\left|
  a^\dagger_{c' mf}a^\dagger_{d' nf'}
  \left(\frac{\lambda_a}{2}\right)_{c' j}
  \left(\frac{\lambda_a}{2}\right)_{d' k}
  \left(\frac{\lambda_b}{2}\right)_{jc}
  \left(\frac{\lambda_b}{2}\right)_{kd}
  a_{dni'}a_{cmi}\right|\hat{\phi_k}\right\rangle \nonumber
 \\*[3mm]
&& 
\phantom{-}\times \delta(\eps_i+\eps_{i'}-\eps_f-\eps_{f'}) 
\sum\limits_{\Sigma\Sigma'}\sum\limits_{qq'}   
\sum\limits_{mm'}   
\frac{g^{\Sigma\Sigma}\,g^{\Sigma'\Sigma'}}        
{4\,\Omega_m^\Sigma \,       
\Omega_{m'}^{\Sigma'}}\,\,       {\cal Q}_{\mbox{\scs
S}}. 
\label{stged}
\eea
Here, we have used the quark and gluon propagators   
in the Feynman gauge \cite{OConnor96}.
The subscripts of the quark creation and annihilation operators stand for 
the color, flavor, and orbital quantum numbers, respectively.
Repeated indices on the Gell-Mann matrices indicate a summation 
according to the Einstein convention.
The quark-gluon vertex 
integrals are combined into the function 
${\cal Q}_{\mbox{\scs S}}$ 
which depends on all the quark and gluon quantum numbers 
involved in the process,
\be \begin{array}{rcclcl}
\D
{\cal Q}_{\mbox{\scs S}}&\D=&\D \phantom{-}&
Q^{m\Sigma}_{fq}\, Q_{f' q'}^{m\Sigma} \,
Q^{m'\Sigma'}_{qi} \,
Q_{q' i'}^{m'\Sigma'}\, E_{{\rm I}}    
&\D -&\D Q^{m\Sigma}_{fq}\, Q_{f' -q'}^{m\Sigma} \,
Q^{m'\Sigma'}_{qi}\, Q_{-q' i'}^{m'\Sigma'}\,  E_{{\rm II}}
\\
&&\D -&\D Q^{m\Sigma}_{f -q}\, Q_{f' q'}^{m\Sigma}\,
   Q^{m'\Sigma'}_{-qi}\,
   Q_{q' i'}^{m'\Sigma'}\, E_{{\rm III}}  
 &\D +&\D Q^{m\Sigma}_{f -q}\,
Q_{f' -q'}^{m\Sigma}\,    
Q^{m'\Sigma'}_{-qi}\,
Q_{-q' i'}^{m'\Sigma'}\, E_{{\rm IV}} .
\end{array}
\label{fourtermsstraight}
\ee
The terms $E_{\rm I}$ to $E_{\rm IV}$, given in Appendix \ref{apendi},
 are sums over the 
energy denominators that arise from the time integrations. 

The energy shift can be interpreted as a two-body operator $\hat V$ in 
second quantization sandwiched between the states $\ket{\phi_k}$. 
This operator
can be translated into first quantization, yielding
\be
V_{12}=\frac{\alpha_{\rm S}^2}{R}\,
 F^2_{12} \,
\sum_\Lambda
\nu_{12}(\Lambda),
\ee
where $\Lambda$ stands for the total exchanged angular momentum 
between the quarks. $F_{12}$ denotes the two-body operator $F_{12}=F_{1}\cdot
F_{2}$.

Expanding the quark-gluon vertex integrals into angular and
radial parts ($S^{m\Sigma}_{nn'}$), we arrive after some algebra at
\begin{eqnarray}
\nu_{12}(\Lambda)&=&
  - \sum\limits_{\Sigma\Sigma'}\sum\limits_{j_q 
   j_{q'}}   \sum\limits_{J_1 J_2} \sum\limits_{\lambda}
   \frac{g^{\Sigma\Sigma}\eta_\Sigma \,
   g^{\Sigma'\Sigma'}\eta_{\Sigma'}}
   {4\, \Omega^\Sigma_mR \, \Omega^{\Sigma'}_{m'}R} 
   (-1)^{J_1+J_2+\Lambda} 
\nonumber\\
&&\times G_{J_1 J_2}(f,q,i)\, G_{J_1 J_2}(f',q',i')\, 
\hat\Lambda^2\,  
   \left\{\begin{array}{ccc}
        j_f & J_1    & j_q\\
        J_2   & j_i  & \Lambda
        \end{array}\right\}
   \left\{\begin{array}{ccc}
        j_{f'} & J_1    & j_{q'}\\
        J_2   &  j_{i'} & \Lambda
        \end{array} \right\} \nonumber\\
&& \times  (-1)^{1+\mu_{f'}-\mu_i} 
     \left(\begin{array}{ccc}
        j_{f} & \Lambda & j_{i}\\
        -\mu_f & \lambda & \mu_i \end{array}\right)
     \left(\begin{array}{ccc}
        j_{f'} & \Lambda & j_{i'}\\
        -\mu_{f'}&-\lambda&\mu_{i'}\end{array}\right) 
       \frac{{\cal S}}{R^3}\ .   \label{437}
\end{eqnarray}
Here, the summation over the radial quantum 
numbers of the intermediate quarks and gluons have  been 
omitted for simplicity. The factors $G$ and $\cal S$ are defined as
\be G_{J_1 J_2}(f,q,i)= \hat\jmath_f \, 
\hat\jmath_i \,  \hat\jmath_{q}^2  \, 
\hat{J}_1\hat{J}_2 \, (-1)^{j_f+j_i-j_{q}}  
\left(\begin{array}{ccc}         j_f & J_1 & j_{q}\\
        \half&0&-\half\end{array}\right)
  \left(\begin{array}{ccc}
        j_{q} & J_2 & j_i\\
        \half&0&-\half\end{array}\right)         
\ee
and
\be\begin{array}{rcclcl}
\D {\cal S}&\D =&\D \phantom{-}& \D
S^{m\Sigma}_{fq}\, S_{f' q'}^{m\Sigma} \,
S^{m'\Sigma'}_{qi} \,
S_{q' i'}^{m'\Sigma'}\, E_{\mbox{\scs I}}    
&\D -&\D S^{m\Sigma}_{fq}\, S_{f' -q'}^{m\Sigma} \,
S^{m'\Sigma'}_{qi}\, S_{-q' i'}^{m'\Sigma'}\,  E_{\mbox{\scs II}}
\\
&&\D -&\D S^{m\Sigma}_{f -q}\, S_{f' q'}^{m\Sigma}\,
   S^{m'\Sigma'}_{-qi}\,
   S_{q' i'}^{m'\Sigma'}\, E_{\mbox{\scs III}}  
 &\D  +&\D S^{m\Sigma}_{f -q}\,
S_{f' -q'}^{m\Sigma}\,    
S^{m'\Sigma'}_{-qi}\,
S_{-q' i'}^{m'\Sigma'}\, E_{\mbox{\scs IV}}\ .
\label{fourS}
\end{array} \ee

Since we are interested in the interactions of quarks and antiquarks 
in the ground-state, we can
restrict ourselves to  $j_n=\half$ with $n=i,i',f,f'$, for which 
Eq.~(\ref{437}) simplifies to
\begin{eqnarray}
\nu_{12}(0)&=& - \sum_{\Sigma\Sigma'}
\sum_{J j_q j_{q'}}
 \frac{g^{\Sigma\Sigma}\eta_\Sigma\,
   g^{\Sigma'\Sigma'}\eta_{\Sigma'}}
   {4\, \Omega^\Sigma_m R \, \Omega^{\Sigma'}_{m'}R} 
  \hat\jmath_q^2\,\hat\jmath_{q'}^2\,\hat{J}^2\,
  \delta_{J_1 J_2}
  \left(\begin{array}{ccc}
        \half & J & j_q\\
        \half&0&-\half\end{array}\right)^2
\nonumber\\*
&&\phantom{- \sum_{\Sigma\Sigma'}
\sum_{J j_q j_{q'}}
 }
  \times \left(\begin{array}{ccc}
        \half & J & j_{q'}\\
        \half&0&-\half\end{array}\right)^2
 \frac{{\cal S}}{R^3},\\*[.5cm]
\nu_{12}(1)&=& -
8\ S^2_{12} 
\sum_{\Sigma\Sigma'}
\sum_{J_1 J_2 j_q j_{q'}}
 \frac{g^{\Sigma\Sigma}\eta_\Sigma \,
   g^{\Sigma'\Sigma'}\eta_{\Sigma'}}
   {4\, \Omega^\Sigma_m R \, \Omega^{\Sigma'}_{m'}R} 
  (-1)^{J_1+J_2-j_q-j_{q'}}\,
  \hat\jmath_q^2 \, \hat\jmath_{q'}^2\,  
  \hat{J}_1^2 \, \hat{J}_2^2
\nonumber\\ 
 && \times \left\{\begin{array}{ccc}
        \half & J_1    & j_q\\
        J_2   & \half & 1
        \end{array}\right\}
   \left\{\begin{array}{ccc}
        \half & J_1    & j_{q'}\\
        J_2   &  \half & 1
        \end{array} \right\} 
    \left(\begin{array}{ccc}
        \half & J_1 & j_q\\
        \half&0&-\half\end{array}\right)
  \label{nu12(1)} 
   \\
&& \times  
  \left(\begin{array}{ccc}
        j_q & J_2 & \half\\
        \half&0&-\half\end{array}\right)         
  \left(\begin{array}{ccc}
        \half & J_1 & j_{q'}\\
        \half&0&-\half\end{array}\right)
  \left(\begin{array}{ccc}
        j_{q'} & J_2 & \half\\
        \half&0&-\half\end{array}\right) \frac{{\cal 
S}}{R^3}\ .         \nonumber 
\end{eqnarray} 
The $\nu_{12}$ can now be determined numerically.
We note that our method differs from that of
ref.~\cite{Stoddart88} in which 
the sum over all intermediate states coupled to exterior states with
$j=0$ or $j=1$ was calculated, while here we evaluate the coefficients 
of the two-body operators directly.


\section{Results and discussion}
The two-body operators for the energy shifts due to the interactions
not discussed above are  may be derived in the same manner. In the
following, we take all the two-body interactions into account.
The two-body operators of the energy shifts can be written as
\begin{mathletters}
\label{two-body}
\be
V_{12}^{\rm (1gx)}=\alphas\,C_{12}^{\rm (1gx)}\,
   \left(A^{\rm (1gx)}+B^{\rm (1gx)}\,S_{12}\right)
\ee
for the one-gluon exchange graph,
\be
V_{12}^{\rm (2gx)}=\alphas^2\,C_{12}^{\rm (2gx)}\,
 \left(A^{\rm (2gx)}+B^{\rm (2gx)}\,S_{12}\right)
\ee
for the straight and crossed two-gluon exchange graphs, and
\be
V_{12}^{\rm (2ga)}=
\alphas^2\,(\mbox{$\frac{1}{4}$}-T_{12})\,C_{12}^{\rm (2ga)}\,
   \left(A^{\rm (2ga)}+B^{\rm (2ga)}\,S_{12}\right)
\ee
for the straight and crossed two-gluon annihilation graphs. The
two-body operators $C_{12}$ and the coefficients $A$ and $B$ are found
in Table \ref{coeffs}.
$V_{12}$ is given in natural units $\hbar c/R$, $R$ being the cavity radius.
$S_{12}$, $T_{12}$ and $F_{12}$ stand for the product of the spin, isospin, and
color operators of the two external particles, respectively.
\end{mathletters}

The one-gluon annihilation graph only contributes to the
energy shifts of quark-antiquark pairs with the quantum-numbers of a
gluon. These do not occur in the hadrons considered here. However, for
completeness we quote the resulting two-body operator as well,
\be
V_{12}^{\rm (1ga)}=0.187505\,{\alpha_{\rm s}}\,(\mbox{$\frac{1}{4}$}-T_{12})
 (F_{12}+\mbox{$\frac{4}{3}$}) (S_{12}+\mbox{$\frac{3}{4}$}).
\ee

At first sight, the results for the two-gluon exchange and
annihilation seem to disagree with those previously obtained  by
Stoddart {\it et al.}\ \cite{Stoddart88}, but since, in contrast to
ref.~\cite{Stoddart88}, we have included all the time-ordered
graphs, this is not surprising. If the coefficients for the two-gluon exchange 
diagrams are calculated to the same maximal energy including   
the time-ordered diagrams calculated in ref.~\cite{Stoddart88} only, 
we are able to reproduce those results. 
For the two-gluon annihilation graphs, we obtain the same result for the 
energy-shift of a quark-antiquark pair in a state $j=0$, even though our 
coefficients  differ from those of ref.~\cite{Stoddart88} for the
unphysical case $j=1$.

\begin{figure}
\hfil
\epsfxsize9cm
\epsfbox{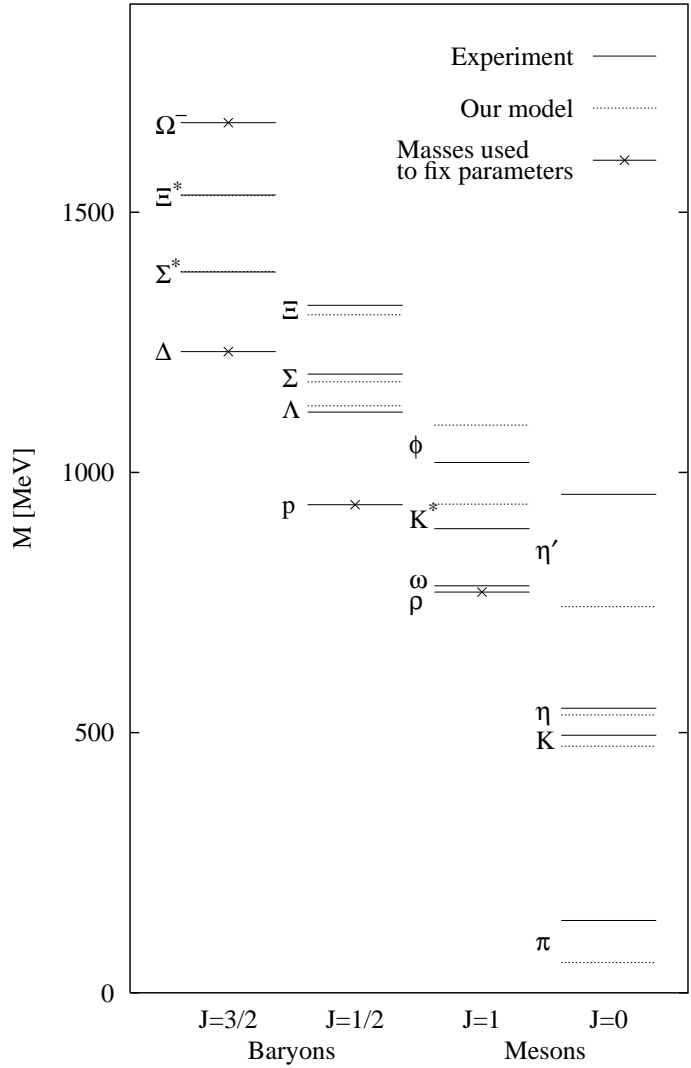}
\hfil

\vspace{0.5cm}
\caption{The hadron spectrum. \label{massspectrum}}
\end{figure}

Using the mass formula of the M.I.T.\ bag model 
\cite{Chodos74a,Chodos74b,DeGrand75},
\be
M=\frac{4}{3}\,(4\pi\,B)^{1/4}\,\Omega^{3/4},
\label{MITmass}
\ee
where $B$ is the bag constant and $\Omega$ is the total energy of the state in 
units of $\hbar c/R$,
we arrive at the mass spectrum for 
the low-lying hadrons (Fig.\ \ref{massspectrum}). 
The total energy $\Omega$ is of the general form
\be
\Omega=N\omega-Z+V_1\alpha_{\rm s}+V_2\alpha_{\rm s}^2,
\label{Omega}
\ee
where $N$ is the number of quarks, $\omega$ the single particle energy,
$Z$ the vacuum energy, $V_1$ and $V_2$ are the first and second order energy
shifts, respectively.  
Consistent with the philosophy mentioned above, we neglect possible
contributions of order $\alphas$ and $\alphas^2$ to $\omega$ and $Z$. 
The masses of the less problematic hadrons, i.e.\ the nucleon, 
the $\Delta$-resonance, 
and the $\rho$-meson, are used to fix the parameters $\alpha_{\rm s}$, $B$ and
$Z$. The $\Omega^{-}$-hyperon  fixes the mass of the strange quark.
The parameters obtained are
\begin{mathletters}
\bea
\alpha_{\rm s} &=& 1.008,\\
Z &=& 1.471,\\
B^{1/4} &=& 158.0\ \mbox{MeV}\\
m_s &=& 1.445\ \mbox{fm}^{-1}=285.1\ \mbox{MeV}.
\eea
\end{mathletters}
The fact that including higher order graphs lowers the value of
$\alpha_{\rm s}$ from 2.2 \cite{Johnson75} in first order to just
over unity in second order is very encouraging.

The two-gluon annihilation graphs lift the degeneracy
of the $\pi$ and the $\eta$ mesons while keeping the $\rho$ and
$\omega$ degenerate. As we are able to
calculate the diagonal and off-diagonal matrix elements due to 
two-gluon annihilation, we may follow the procedure outlined by Donoghue
and Gomm \cite{Donoghue83}  and diagonalize the Hamilton
matrix, which in the basis $(u\bar u+ d\bar d)/\sqrt{2}$ and $s\bar s$
reads
\be
\hat \Omega = \left(\begin{array}{cc} \Omega_{11} & \Omega_{12} \\ 
				\Omega_{21} & \Omega_{22}
			\end{array}\right),
\ee
with
\begin{mathletters}
\bea
\Omega_{11} &=& 2\omega_{u} -Z + \alphas\ V_{1}(u\to u) + \alphas^2\
\left(V_{2,e}(u\to u)+2\,V_{2,a}(u\to u)\right)\ , \\
\Omega_{12} &=&  \sqrt{2}\,\alphas^2\ V_{2,a}(u\to s)\ ,\\
\Omega_{21} &=&  \sqrt{2}\,\alphas^2\ V_{2,a}(s\to u)\ ,\\
\Omega_{22} &=& 2\omega_{s} -Z + \alphas\ V_{1}(s\to s) + \alphas^2\
\left(V_{2,e}(s\to s)+V_{2,a}(s\to s)\right)\ .
\eea
\end{mathletters}
Here, $u$ denotes a massless quark, $s$ a strange quark, and the
labels
$e$ and $a$ refer to interactions due to gluon exchange and
annihilation. 
We differ slightly from the procedure of Donoghue and Gomm
\cite{Donoghue83}, as
they use the {\em experimental} values of the pion and kaon masses in their
calculation. 
Deriving the masses of $\eta$ and $\eta'$ in the
framework of the M.I.T.\ bag model by diagonalizing the submatrix 
$\hat \Omega$ and
inserting the eigenvalues into Eq.~(\ref{MITmass}), we obtain
\begin{mathletters}
\bea
m_\eta &=& 534\ \mbox{MeV}, \\
m_{\eta'} &=& 742\ \mbox{MeV},
\eea
\end{mathletters}
with a mixing angle of $33^\circ$.
The resulting $\eta$ mass is very close to the experimental value
of $m_\eta=547\ \mbox{MeV}$, while the  $\eta'$ mass is too low
compared with $m_{\eta'}=958\ \mbox{MeV}$. However, here we have only taken
into account contributions from massless and strange quarks to the mass of the
$\eta'$. The $\eta'$ meson could contain a significant $c\bar c$
component which might well account 
for the remaining difference, since heavy quark masses have an
important influence on the energy eigenvalues.

As one can see in Fig.\ \ref{massspectrum}, the 
fit is very good for 
the masses of the baryons. The splitting between the $\Lambda$ and the $\Sigma$
is much better than in the calculation to first order in $\alpha_{\rm s}$
\cite {Johnson75}. However, the predictions for the vector mesons $\phi$ 
and $K^\ast$ are still
too high, while for the pseudoscalar mesons, $\pi$, $K$, $\eta$ and
$\eta'$, they are too low. 

\begin{figure}
\begin{picture}(37000,10000)
\drawline\gluon[\E\CURLY](1000,5000)[4]
\put(\pmidx,0){\makebox(0,0){(1gx)}}
\drawline\fermion[\N\REG](\gluonfrontx,\gluonfronty)[\gluonlengthx]
\drawline\fermion[\N\REG](\gluonbackx,\gluonbacky)[\gluonlengthx]
\drawline\fermion[\S\REG](\gluonfrontx,\gluonfronty)[\gluonlengthx]
\drawline\fermion[\S\REG](\gluonbackx,\gluonbacky)[\gluonlengthx]
\drawline\gluon[\S\CURLY](9000,6700)[4]
\put(\pmidx,0){\makebox(0,0){(1ga)}}
\setlength{\fermilength}{0.707\gluonlengthy}
\drawline\fermion[\SE\REG](\gluonbackx,\gluonbacky)[\fermilength]
\drawline\fermion[\SW\REG](\gluonbackx,\gluonbacky)[\fermilength]
\drawline\fermion[\NE\REG](\gluonfrontx,\gluonfronty)[\fermilength]
\drawline\fermion[\NW\REG](\gluonfrontx,\gluonfronty)[\fermilength]
\drawline\gluon[\E\CURLY](14000,3850)[4] 
\put(\pmidx,0){\makebox(0,0){(s2gx)}} 
\setlength{\fermilength}{0.666\gluonlengthx} 
\setlength{\gluonfourlength}{1\gluonlengthx} 
\drawline\fermion[\N\REG](\gluonbackx,\gluonbacky)[\fermilength] 
\drawline\fermion[\S\REG](\gluonbackx,\gluonbacky)[\fermilength] 
\drawline\fermion[\S\REG](\gluonfrontx,\gluonfronty)[\fermilength] 
\drawline\fermion[\N\REG](\gluonfrontx,\gluonfronty)[\fermilength] 
\drawline\gluon[\E\CURLY](\fermionbackx,\fermionbacky)[4] 
\drawline\fermion[\N\REG](\gluonbackx,\gluonbacky)[\fermilength] 
\drawline\fermion[\S\REG](\gluonbackx,\gluonbacky)[\fermilength] 
\drawline\fermion[\S\REG](\gluonfrontx,\gluonfronty)[\fermilength] 
\drawline\fermion[\N\REG](\gluonfrontx,\gluonfronty)[\fermilength] 
\drawline\gluon[\NE\REG](20500,3080)[3] 
\put(\pmidx,0){\makebox(0,0){(c2gx)}} 
\setlength{\templength}{2\gluonfourlength} 
\addtolength{\templength}{-1\gluonlengthy} 
\setlength{\fermilength}{0.49\templength} 
\drawline\fermion[\S\REG](\gluonfrontx,\gluonfronty)[\fermilength] 
\drawline\fermion[\S\REG](\gluonbackx,\gluonbacky)[\gluonlengthy] 
\drawline\fermion[\N\REG](\gluonbackx,\gluonbacky)[\fermilength] 
\drawline\fermion[\N\REG](\gluonfrontx,\gluonfronty)[\gluonlengthy] 
\drawline\gluon[\SE\REG](\fermionbackx,\fermionbacky)[3] 
\drawline\fermion[\N\REG](\gluonfrontx,\gluonfronty)[\fermilength] 
\drawline\fermion[\S\REG](\gluonbackx,\gluonbacky)[\fermilength] 
\drawline\gluon[\N\CURLY](27000,3300)[4] 
\setlength{\fermilength}{0.49\gluonlengthy} 
\drawline\fermion[\N\REG](\gluonbackx,\gluonbacky)[\fermilength] 
\drawline\fermion[\S\REG](\gluonfrontx,\gluonfronty)[\fermilength] 
\drawline\fermion[\E\REG](\gluonbackx,\gluonbacky)[\gluonlengthy] 
\put(\pmidx,0){\makebox(0,0){(s2ga)}} 
\drawline\gluon[\S\CURLY](\pbackx,\pbacky)[4] 
\drawline\fermion[\S\REG](\gluonbackx,\gluonbacky)[\fermilength] 
\drawline\fermion[\N\REG](\gluonfrontx,\gluonfronty)[\fermilength] 
\drawline\fermion[\W\REG](\gluonbackx,\gluonbacky)[\gluonlengthy] 
\drawline\gluon[\NE\REG](33000,3100)[3] 
\put(\pmidx,0){\makebox(0,0){(c2ga)}} 
\setlength{\templength}{2\gluonfourlength} 
\addtolength{\templength}{-1\gluonlengthy} 
\setlength{\fermilength}{0.49\templength} 
\drawline\fermion[\N\REG](\gluonbackx,\gluonbacky)[\fermilength] 
\drawline\fermion[\S\REG](\gluonfrontx,\gluonfronty)[\fermilength] 
\drawline\fermion[\E\REG](\gluonfrontx,\gluonfronty)[\gluonlengthx] 
\drawline\fermion[\W\REG](\gluonbackx,\gluonbacky)[\gluonlengthx] 
\drawline\gluon[\SE\REG](\pbackx,\pbacky)[3] 
\drawline\fermion[\N\REG](\gluonfrontx,\gluonfronty)[\fermilength] 
\drawline\fermion[\S\REG](\gluonbackx,\gluonbacky)[\fermilength] 
\end{picture} 
\vspace{0.4cm}
\caption{Finite Feynman diagrams of first and second order in $\alpha_{\rm s}$. 
\label{Feynman}} 
\end{figure}
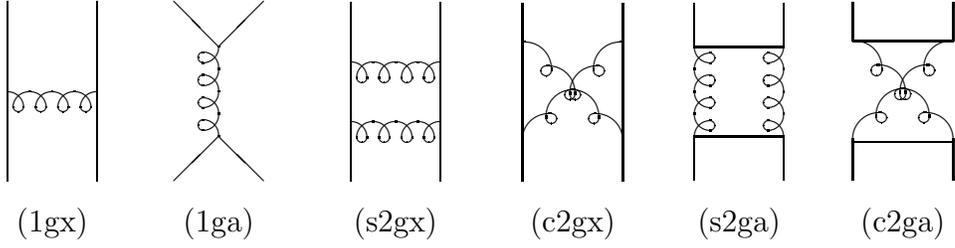

The fit only takes into account the two-body diagrams
calculated in this paper, but
none of the divergent 
graphs to order $\alpha_{\rm s}^2$, i.e.\ those that have a loop 
on any of the external or internal lines or vertices in Figs.\ \ref{Feynman}(1gx) 
and \ref{Feynman}(1ga). These divergent
graphs should be included to make the second order energy shift gauge invariant. 
However, even though the cavity renormalization techniques have been
developed \cite{Page93,Schreiber92b,Stoddart90,Cuthbert93}, the
numerical effort required to calculate such graphs is 
considerable\footnote{The regularization procedure relies on delicate 
cancellations between oscillating terms.} 
and
beyond the scope of this paper. 
In any case, these contributions are of the same form as the  two-body 
operators for one-gluon exchange and annihilation. 
They will merely renormalize the first-order contribution by a factor. 
Thus the part which is orthogonal to the first-order contribution will
still be gauge independent. The fit has been found without the introduction of
this factor.

\section{Conclusion}
We have calculated the non-diverging Feynman diagrams up to second
order in $\alphas$ in the framework of cavity quantum chromodynamics
and obtained the two-body operators for the energy shifts. The free
parameters of the model, the strong coupling constant $\alpha_s$, the
zero-point energy $Z_0$, the bag pressure $B$, and the mass of the
strange quark $m_s$, are fixed by four out of the sixteen available
masses of the hadrons containing only  up, down, and strange quarks
coupled to total angular momentum of up to $3/2$.
The calculated masses agree very well with the experimental data. It is
especially noteworthy, that the mass of the $\eta$-meson can be
calculated by diagonalizing the $\eta$--$\eta'$ subspace Hamiltonian
without introducing any further parameters into the model.

The inclusion of second order interactions leads to a much smaller value of
$\alphas$ ($\approx 1$) than only considering first order terms
$(\alphas>2)$.
Self-energies will be included in future work, but at the moment we
neglect them as they might be taken care of by the boundary conditions
already.

There is still a lot of work ahead, to complete the calculations in cavity QCD
to order $\alphas^2$. This includes the evaluation of divergent diagrams, 
but also the calculation of other observables than the masses, e.g.\ the 
magnetic moments, charge radii, and the ratio $g_A/g_V$.

\acknowledgements
This work is supported in part by the Foundation for Fundamental
Research (FFR).

\appendix
\section{``Straight'' two-gluon exchange energy-denominators}
\label{apendi}
\be
\begin{array}{rcl}
E_{\mbox{\scs I}} &=& \phantom{+}\left[(\eqp-\eip+\omsp)(\efp-\eip+\oms+\omsp)
                            (\eq-\ef+\oms)\right]^{-1}\\
             &&+\left[(\eqp-\eip+\omsp)(-\ei-\eip+\eq+\eqp)
                      (\eq-\ef+\oms)\right]^{-1}\\
             &&+\left[(\eq-\ei+\omsp)(-\ei-\eip+\eq+\eqp)
                      (\eq-\ef+\oms)\right]^{-1}\\
             &&+\left[(\eqp-\eip+\omsp)(-\ei-\eip+\eq+\eqp)
                      (\eqp-\efp+\oms)\right]^{-1}\\
         &&+\left[(\eq-\ei+\omsp)(-\ei-\eip+\eq+\eqp)
                  (\eqp-\efp+\oms)\right]^{-1}\\
         &&+\left[(\eq-\ei+\omsp)(\ef-\ei+\oms+\omsp)
                  (\eqp-\efp+\oms)\right]^{-1}
\end{array}
\ee

\be
\begin{array}{rcl}
E_{\mbox{\scs II}} &=&\phantom{+}\left[(\eqp+\efp+\oms)(\efp-\eip+\oms+\omsp)
                        (\eq-\ef+\oms)\right]^{-1}\\
              &&+\left[(\eqp+\efp+\oms)(\efp-\ei+\eq+\eqp+\oms+\omsp)
                       (\eq-\ef+\oms)\right]^{-1}\\
              &&+\left[(\eq-\ei+\omsp)(\efp-\ei+\eq+\eqp+\oms+\omsp)
                       (\eq-\ef+\oms)\right]^{-1}\\
          &&+\left[(\eqp+\efp+\oms)(\efp-\ei+\eq+\eqp+\oms+\omsp)
                       (\eqp+\eip+\omsp)\right]^{-1}\\
              &&+\left[(\eq-\ei+\omsp)(\efp-\ei+\eq+\eqp+\oms+\omsp)
                       (\eqp+\eip+\omsp)\right]^{-1}\\
              &&+\left[(\eq-\ei+\omsp)(\ef-\ei+\oms+\omsp)
                       (\eqp+\eip+\omsp)\right]^{-1}
\end{array}\ee

\be\begin{array}{rcl}
E_{\mbox{\scs III}} &=&\phantom{+}\left[(\eqp-\eip+\omsp)(\efp-\eip+\oms+\omsp)
                             (\eq+\ei+\omsp)\right]^{-1}\\ 
                    &&+\left[(\eqp-\eip+\omsp)(\ef-\eip+\eq+\eqp+\oms+\omsp)
                             (\eq+\ei+\omsp)\right]^{-1}\\     
                    &&+\left[(\eqp-\eip+\omsp)(\ef-\eip+\eq+\eqp+\oms+\omsp)
                             (\eqp-\efp+\oms)\right]^{-1}\\     
                    &&+\left[(\eq+\ef+\oms)(\ef-\eip+\eq+\eqp+\oms+\omsp)
                             (\eq+\ei+\omsp)\right]^{-1}\\  
                    &&+\left[(\eq+\ef+\oms)(\ef-\eip+\eq+\eqp+\oms+\omsp)
                             (\eqp-\efp+\oms)\right]^{-1}\\     
                    &&+\left[(\eq+\ef+\oms)(\ef-\ei+\oms+\omsp)
                             (\eqp-\efp+\oms)\right]^{-1}
\end{array}\ee

\be\begin{array}{rcl}
E_{\mbox{\scs IV}} &=&\phantom{+}\left[(\eqp+\efp+\oms)(\efp-\eip+\oms+\omsp)
                            (\eq+\ei+\omsp)\right]^{-1}\\ 
            &&+\left[(\eqp+\efp+\oms)(\ef+\efp+\eq+\eqp)
                            (\eq+\ei+\omsp)\right]^{-1}\\ 
                    &&+\left[(\eq+\ef+\oms)(\ef+\efp+\eq+\eqp)
                            (\eq+\ei+\omsp)\right]^{-1}\\ 
                    &&+\left[(\eqp+\efp+\oms)(\ef+\efp+\eq+\eqp)
                            (\eqp+\eip+\omsp)\right]^{-1}\\ 
                    &&+\left[(\eq+\ef+\oms)(\ef+\efp+\eq+\eqp)
                            (\eqp+\eip+\omsp)\right]^{-1}\\ 
                    &&+\left[(\eq+\ef+\oms)(\ef-\ei+\oms+\omsp)
                            (\eqp+\eip+\omsp)\right]^{-1}
\end{array}
\ee


\begin{table}
\squeezetable
\begin{tabular}{lcddddd}
&& \multicolumn{1}{c}{\bf 1gx} & \multicolumn{1}{c}{\bf s2gx} & 
\multicolumn{1}{c}{\bf c2gx} & \multicolumn{1}{c}{\bf s2ga} & 
\multicolumn{1}{c}{\bf c2ga} \\ \tableline
& $C_{12}$ & \multicolumn{1}{c}{$F_{12}$}
& \multicolumn{1}{c}{$F_{12}^2$}
& \multicolumn{1}{c}{$F_{12}\left(F_{12}+\frac{3}{2}\right)$}
& \multicolumn{1}{c}{$2
\left(\frac{16}{27}-\frac{1}{18}F_{12}\right)$}
&  \multicolumn{1}{c}{$2
\left(-\frac{2}{27}-\frac{5}{9}F_{12}\right)$}
\\ \tableline
$m_1=0$ & $A$ & 0.009795 & $-$0.406119 & 0.146124 & 0.077642 &
0.286519\\
$m_2=0$	& $B$      & $-$0.708080  &    0.747443 & $-$0.190998 & 0.003926 &
	$-$0.331505\\ \tableline
$m_1=0$ & $A$ & 0.022668 & $-$0.398787 & 0.105459 & 0.020012 & 0.094710 \\
$m_2=m_s$ & $B$ & $-$0.566556 & 0.613761 & $-$0.074720 & $-$0.015174 &
$-$0.092402\\ \tableline
$m_1=m_s$ & $A$ & 0.052685 & $-$0.422564 & 0.067182 & $-$0.037616 &
$-$0.084483 \\
$m_2=m_s$ & $B$ & $-$0.457013 & 0.551227 &0.009522 & $-$0.017055 &
0.036234\\
\end{tabular}
\vspace{2ex}
\caption{\label{coeffs} The numerical results for the coefficients of the two-body
operators as defined in Eqs.\ (\ref{two-body}). The abbreviations stand for one-gluon
exchange, straight two-gluon exchange, crossed two-gluon exchange,
straight two-gluon annihilation, and crossed two-gluon annihilation,
respectively. The second row gives the appropriate color factors.} 
\end{table}


\begin{thebibliography}{10}

\bibitem{Chodos74a}
A. Chodos {\it et~al.}, Phys. Rev. D {\bf 9} (1974)  3471.

\bibitem{Chodos74b}
A. Chodos, R.~L. Jaffe, K. Johnson, and C.~B. Thorn, Phys. Rev. D {\bf 10}  (1974)  2599.

\bibitem{DeGrand75}
T. DeGrand, R.~L. Jaffe, K. Johnson, and J. Kiskis, Phys. Rev. D {\bf 12}  (1975)
  2060.

\bibitem{Buser88}
R.~F. Buser, R.~D. Viollier, and P. Zimak, Int. J. Theo. Phys. {\bf 27}  (1988)
 925.

\bibitem{Page93}
P.~R. Page, R.~J. Lindebaum, and R.~D. Viollier, Nucl. Phys. A {\bf 560}   (1993)
  1003.

\bibitem{Schreiber92b}
G.~U. Schreiber and R.~D. Viollier, Phys. Lett. B {\bf 279} (1992)  131.

\bibitem{Stoddart90}
A.~J. Stoddart and R.~D. Viollier, Phys. Lett. B {\bf 236}  (1990) 387.

\bibitem{Cuthbert93}
J.~A. Cuthbert and R.~D. Viollier, Z. Phys. C {\bf 58}  (1993) 295.

\bibitem{Sucher57}
J. Sucher, Phys. Rev. {\bf 107}  (1957)  1448.

\bibitem{Stoddart88}
A.~J. Stoddart and R.~D. Viollier, Phys. Lett. B {\bf 208} (1988)  65.

\bibitem{OConnor96}
M.~S. O'Connor and R.~D. Viollier, Ann. Phys. (N.Y.) {\bf 248}  (1996)  286.

\bibitem{Johnson75}
K. Johnson, Acta Physica Polonica B {\bf 6}  (1975)  865.

\bibitem{Donoghue83}
J.~F. Donoghue and H. Gomm, Phys. Rev. D {\bf 28}  (1983)  2800.

\end{thebibliography}
\end{document}